\documentclass[%
 reprint,
superscriptaddress,
 amsmath,amssymb,
 aps,
]{revtex4-1}

\usepackage{graphicx}
\usepackage{dcolumn}
\usepackage{bm}
\usepackage{hyperref}
\usepackage[mathlines]{lineno}
\usepackage{multirow}
\usepackage{array}

\usepackage{color}

\begin{document}

\preprint{APS/123-QED}

\title{First search for Lorentz and CPT violation in double beta decay\\with EXO-200}

\newcommand{\Alabama}{\affiliation{Department of Physics and Astronomy, University of Alabama, Tuscaloosa, Alabama 35487, USA}}
\newcommand{\Alberta}{\affiliation{University of Alberta, Edmonton, Alberta, Canada}}
\newcommand{\Bern}{\affiliation{LHEP, Albert Einstein Center, University of Bern, Bern, Switzerland}}
\newcommand{\CALTECH}{\affiliation{Kellogg Lab, Caltech, Pasadena, California 91125, USA}}
\newcommand{\Carleton}{\affiliation{Physics Department, Carleton University, Ottawa, Ontario K1S 5B6, Canada}}
\newcommand{\CSU}{\affiliation{Physics Department, Colorado State University, Fort Collins, Colorado 80523, USA}}
\newcommand{\Drexel}{\affiliation{Department of Physics, Drexel University, Philadelphia, Pennsylvania 19104, USA}}
\newcommand{\Duke}{\affiliation{Department of Physics, Duke University, and Triangle Universities Nuclear Laboratory (TUNL), Durham, North Carolina 27708, USA}}
\newcommand{\IBS}{\affiliation{IBS Center for Underground Physics, Daejeon, Korea}}
\newcommand{\IHEP}{\affiliation{Institute of High Energy Physics, Beijing, China}}
\newcommand{\Illinois}{\affiliation{Physics Department, University of Illinois, Urbana-Champaign, Illinois 61801, USA}}
\newcommand{\Indiana}{\affiliation{Physics Department and CEEM, Indiana University, Bloomington, Indiana 47405, USA}}
\newcommand{\ITEP}{\affiliation{Institute for Theoretical and Experimental Physics, Moscow, Russia}}
\newcommand{\Laurentian}{\affiliation{Department of Physics, Laurentian University, Sudbury, Ontario P3E 2C6, Canada}}
\newcommand{\Maryland}{\affiliation{Physics Department, University of Maryland, College Park, Maryland 20742, USA}}
\newcommand{\Munich}{\affiliation{Technische Universit\"at M\"unchen, Physikdepartment and Excellence Cluster Universe, Garching 80805, Germany}}
\newcommand{\SDakota}{\affiliation{Physics Department, University of South Dakota, Vermillion, South Dakota 57069, USA}}
\newcommand{\Seoul}{\affiliation{Department of Physics, University of Seoul, Seoul, Korea}}
\newcommand{\SLAC}{\affiliation{SLAC National Accelerator Laboratory, Stanford, California 94025, USA}}
\newcommand{\Stanford}{\affiliation{Physics Department, Stanford University, Stanford, California 94305, USA}}
\newcommand{\Stony}{\affiliation{Department of Physics and Astronomy, Stony Brook University, SUNY, Stony Brook, New York 11794, USA}}
\newcommand{\TRIUMF}{\affiliation{TRIUMF, Vancouver, BC, Canada}}
\newcommand{\UMass}{\affiliation{Amherst Center for Fundamental Interactions and Physics Department, University of Massachusetts, Amherst, MA 01003, USA}}
\newcommand{\WIPP}{\affiliation{Waste Isolation Pilot Plant, Carlsbad, New Mexico 88220, USA}}
\newcommand{\Caltech}{\affiliation{Kellogg Lab, Caltech, Pasadena, CA, USA}}
\newcommand{\Karlsruhe}{\affiliation{Institute for Theoretical Physics, Karlsruhe Institute of Technology, Karlsruhe, Germany}}
\newcommand{\McGill}{\affiliation{Physics Department, McGill University, Montreal, Quebec H3A 2T8, Canada}}
\author{J.B.~Albert}\Indiana
\author{P.S.~Barbeau}\Duke
\author{D.~Beck}\Illinois
\author{V.~Belov}\ITEP
\author{M.~Breidenbach}\SLAC
\author{T.~Brunner}\TRIUMF\McGill
\author{A.~Burenkov}\ITEP
\author{G.F.~Cao}\IHEP
\author{C.~Chambers}\CSU
\author{B.~Cleveland}\altaffiliation{Also SNOLAB, Sudbury, ON, Canada}\Laurentian
\author{M.~Coon}\Illinois
\author{A.~Craycraft}\CSU
\author{T.~Daniels}\SLAC
\author{M.~Danilov}\ITEP
\author{S.J.~Daugherty}\Indiana
\author{C.G.~Davis}\altaffiliation{Present address: Naval Research Lab, Washington D.C., USA}\Maryland
\author{J.~Davis}\SLAC
\author{S.~Delaquis}\Bern
\author{A.~Der Mesrobian-Kabakian}\Laurentian
\author{R.~DeVoe}\Stanford
\author{J.S.~D\'iaz}\Karlsruhe
\author{T.~Didberidze}\Alabama
\author{J.~Dilling}\TRIUMF
\author{A.~Dolgolenko}\ITEP
\author{M.J.~Dolinski}\Drexel
\author{M.~Dunford}\Carleton
\author{W.~Fairbank Jr.}\CSU
\author{J.~Farine}\Laurentian
\author{S.~Feyzbkhsh}\UMass
\author{W.~Feldmeier}\Munich
\author{P.~Fierlinger}\Munich
\author{D.~Fudenberg}\Stanford
\author{R.~Gornea}\Carleton
\author{K.~Graham}\Carleton
\author{G.~Gratta}\Stanford
\author{C.~Hall}\Maryland
\author{S.~Homiller}\Illinois
\author{M.~Hughes}\Alabama
\author{M.J.~Jewell}\Stanford
\author{X.S.~Jiang}\IHEP
\author{A.~Johnson}\SLAC
\author{T.N.~Johnson}\email[Corresponding author: ]{tesjohns@ucdavis.edu}\altaffiliation{Present address: University of California, Davis, CA  95616}\Indiana 
\author{S.~Johnston}\UMass
\author{A.~Karelin}\ITEP
\author{L.J.~Kaufman}\Indiana
\author{R.~Killick}\Carleton
\author{T.~Koffas}\Carleton
\author{S.~Kravitz}\Stanford
\author{R.~Kr\"ucken}\TRIUMF
\author{A.~Kuchenkov}\ITEP
\author{K.S.~Kumar}\Stony
\author{D.S.~Leonard}\IBS
\author{C.~Licciardi}\Carleton
\author{Y.H.~Lin}\Drexel
\author{J.~Ling}\altaffiliation{Present address: Sun Yat-Sen University, Guangzhou 510275, China}\Illinois
\author{R.~MacLellan}\SDakota
\author{M.G.~Marino}\Munich
\author{B.~Mong}\Laurentian
\author{D.~Moore}\Stanford
\author{R.~Nelson}\WIPP
\author{O.~Njoya}\Stony
\author{A.~Odian}\SLAC
\author{I.~Ostrovskiy}\Stanford
\author{A.~Piepke}\Alabama
\author{A.~Pocar}\UMass
\author{C.Y.~Prescott}\SLAC
\author{F. Reti\'ere}\TRIUMF
\author{P.C.~Rowson}\SLAC
\author{J.J.~Russell}\SLAC
\author{A.~Schubert}\Stanford
\author{D.~Sinclair}\TRIUMF\Carleton
\author{E.~Smith}\Drexel
\author{V.~Stekhanov}\ITEP
\author{M.~Tarka}\Stony
\author{T.~Tolba}\altaffiliation{Present address: Institute of High Energy Physics, Beijing, China}\Bern
\author{R.~Tsang}\Alabama
\author{K.~Twelker}\Stanford
\author{J.-L.~Vuilleumier}\Bern
\author{P.~Vogel}\Caltech
\author{A.~Waite}\SLAC
\author{J.~Walton}\Illinois
\author{T.~Walton}\CSU
\author{M.~Weber}\Stanford
\author{L.J.~Wen}\IHEP
\author{U.~Wichoski}\Laurentian
\author{J.~Wood}\WIPP
\author{L.~Yang}\Illinois
\author{Y.-R.~Yen}\Drexel
\author{O.Ya.~Zeldovich}\ITEP

\collaboration{EXO-200 Collaboration}

\date{\today}

\begin{abstract}
A search for Lorentz- and CPT-violating signals in the double beta decay spectrum of $^{136}$Xe has been performed using an exposure of 100 kg$\cdot$yr with the EXO-200 detector. No significant evidence of the spectral modification due to isotropic Lorentz-violation was found, and a two-sided limit of $-2.65 \times 10^{-5 } \; \textrm{GeV} < \mathring{a}^{(3)}_{\text{of}} < 7.60 \times 10^{-6} \;  \textrm{GeV}$ (90\% C.L.)  is placed on the relevant coefficient within the Standard-Model Extension (SME). This is the first experimental study of the effect of the SME-defined oscillation-free and momentum-independent neutrino coupling operator on the double beta decay process.

\end{abstract}

\maketitle

\section{\label{Introduction}Introduction}

A central goal of physics is the development of a consistent framework unifying quantum mechanics and general relativity.  Different approaches have been implemented to reconcile these two successful descriptions of nature.  In this process, it was discovered that the breakdown of Lorentz and CPT (the combination of Charge, Parity, and Time-reversal transformations) symmetries at the Planck scale could arise in many candidates for a description of quantum gravity \cite{K_OGpaper,Kost1995,Carroll2001,Jackiw2003,Mavromatos2005,Amelino2013}.  The standard model of particle physics (SM), which has with few exceptions remained experimentally robust, assumes complete invariance under Lorentz transformations (boosts and rotations) which leads to invariance under CPT transformation.  The observation of the violation of either of these symmetries would imply the observation of new physics beyond the SM.

Direct observation of physics at the Planck scale is not yet possible (length scales of $\sim10^{-35}$ m, and energy scales of $\sim10^{19}$ GeV.)  However, it is possible that unconventional physics at very high energies could lead to suppressed effects at low energies, potentially observable with current experimental technologies.  As almost all physical measurements that have been made to date have been compatible with the SM, a good candidate theory would be one that reduces to the SM at a particular limit - specifically at the limits that we have been able to probe by experiment.  This theory could include Lorentz-violating operators which remain unobserved because their effects couple to quantities that are challenging to measure, such as gravity or weak interactions \cite{Tasson2009,BetaDecays_LV_DKL,Dijck:2013dza}.

The Standard-Model Extension (SME), developed by Kosteleck\'{y} et al.~\cite{SME1, SME2, SME3},  provides a framework that meets these experimental requirements.  In flat spacetime, the SME describes the interaction between SM particles with uniform and constant tensor fields permeating all of spacetime. These background fields would arise as non-zero vacuum expectation values of dynamical fields in the underlying theory.  The coupling between SM particles and these background fields is characterized by so-called SME coefficients, which control the size of the breakdown of Lorentz symmetry.  Each coefficient would need to be determined by experimental observation of the effect of the tensor field on particle interactions.  The potential effects of these couplings on observable physics have been described for many sectors of physics, and experimental limits have been set on hundreds of SME coefficients.  The current limits are compiled into a data table that is updated with new results annually \cite{K_DataTables_LV}. 

Neutrinos are an especially interesting probe of unconventional physics because they mainly interact through the weak interaction, which has been shown to break symmetries previously thought to be exact \cite{Wu, Cronin}.  The operators that couple to neutrinos in the SME affect the flavor oscillation properties, neutrino velocity, or phase space, often revealed as sidereal time dependence \cite{Mewes2004,LVnuReview}.  Many experiments have searched directly for these effects and set limits on the coupling coefficients related to these particular tensor fields \cite{DC_LV_1, MB_LV_1, MB_LV_2, IC_LV_1, LSND_LV_1, MN_LV_1, MN_LV_2, MN_LV_3,LV_SK,Diaz:2016fqd}.  Other conservative limits have been estimated by analyzing published experimental data \cite{ LV_RebelMufson,KM_NeutrinosAndLV, BetaDecays_LV_DKL}.

To date, most of the direct searches for deviations from exact Lorentz and CPT invariance in the SME framework using neutrinos \cite{DC_LV_1, MB_LV_1, MB_LV_2, IC_LV_1, LSND_LV_1, MN_LV_1, MN_LV_2, MN_LV_3,LV_SK,Diaz:2016fqd} have been based on oscillation.  Nonetheless, some Lorentz-violating effects could remain undetected because of the nature of the corresponding experimental signatures.  There exists an operator in the SME that couples to neutrinos which is momentum-independent and does not affect neutrino oscillation (oscillation-free) and is unobservable to long-baseline neutrino experiments \cite{BetaDecays_LV_DKL}.  This renormalizable Lorentz-violating operator, known as the {\it countershaded} operator, has mass dimension three and also breaks CPT.  The corresponding countershaded coefficient has four components, one time-like ($a^{(3)}_{\text{of}})_{00}$ and three space-like ($a^{(3)}_{\text{of}})_{1m}$, with $m=0,\pm1$.  A non-zero value of ($a^{(3)}_{\text{of}})_{00}$ would produce small deviations in the shape of the energy spectrum for single or double beta decay, which can be searched for experimentally.  In order to probe the space-like components, measurement of the direction of the emitted betas is required.  In this work we employ a new method to explore for the first time the effects of the countershaded coefficient $(a^{(3)}_{\text{of}})_{00}$ on a wide energy range of the double beta decay spectrum.

\section{Detector Description}

The EXO-200 detector was built to measure the two-neutrino double beta decay ($2\nu\beta\beta$) spectrum of $^{136}$Xe and to search for the neutrinoless version ($0\nu\beta\beta$) by measuring the electron energy sum spectrum from these processes with high precision.  EXO-200 is a good candidate detector to search for the effects of the time-like component on double beta decay due to the low background of the experiment and ability to precisely measure the spectral shape.

The EXO-200 detector is described in detail elsewhere \cite{Auger:2012gs}.  In summary, the detector is made up of two back to back cylindrical time projection chambers (TPCs) that share a central cathode, each roughly 40 cm in diameter and 22 cm in length.  The detector is filled with liquid xenon (LXe) that has been enriched to 80.6\% $^{136}$Xe.  The LXe is continually circulated through purifiers to remove electronegative impurities. 

Ionizing particle interactions in the LXe produce both scintillation light and ionization electrons.  The scintillation light is reflected by a teflon shell around the barrel of the detector and collected by an array of large area avalanche photodiodes \cite{APDs} at the end-cap of each TPC.  The free electrons are drifted by an electric field toward the end-caps of the TPC, passing a shielding wire plane (``v-plane'') on which signals are induced.  The charge is then collected on a wire plane that acts as the anode (``u-plane'') held at virtual ground, which is crossed at 60$^{\circ}$ from the v-plane.  Signals are grouped together into ``clusters'' which correspond to a single, localized energy deposition in the detector.  The combination of a scintillation signal with both a u- and v- signal allows the position of the cluster to be fully determined.

The detector is held in a bath of HFE-7000 cryogenic fluid, which is contained by a double-walled, vacuum insulated copper cryostat.  The cryostat is surrounded by a 25 cm thick lead shield.  The detector is mounted to the cryostat by copper legs, which also serve as conduits for electronics wiring and LXe flow.

The detector system is located in a clean room under an overburden of 1624 m.w.e.~\cite{EXO200::2015wtc} at the Waste Isolation Pilot Plant mine near Carlsbad, NM, USA.  It is surrounded by an active muon veto system, which identifies 96\%~\cite{EXO200::2015wtc} of muons passing though the TPC and allows rejection of prompt cosmogenic backgrounds.

\section{Analysis Method}

The coupling of a neutrino to the countershaded operator alters the neutrino momentum from the standard $q^{\alpha} = ( \omega, \textbf{\textit{q}})$ to $q^{\alpha} = ( \omega, \textbf{\textit{q}} + \textbf{\textit{a}}^{(3)}_{\text{of}} - \mathring{a}^{(3)}_{\text{of}} \hat{\textbf{\textit{q}}})$ \cite{DBD_LV_JD}.  This deviation in the neutrino momentum modifies the double beta decay transition amplitude as well as the neutrino dispersion relation \cite{DBD_LV_JD}.  This leads to a differential decay rate in terms of the kinetic energies of the two emitted electrons (all energy variables are given in terms of $m_e$) of
\begin{align} \label{eq:DiffDecayRate}
\nonumber
\frac{d\Gamma}{dt_1dt_2} = & \left[ \frac{G^4_F |V_{ud}|^4 g^4_A m^{11}_e}{240 \pi ^7} |M^{2\nu}|^2 \right]F(Z,t_1)F(Z,t_2) \\ & \times |\textbf{\textit{p}}_1| (t_1 + 1) |\textbf{\textit{p}}_2| (t_2 + 1) (\hat{\omega}^5_0 + 10 \mathring{a}^{(3)}_{\text{of}} \hat{\omega}^4_0 ),
\end{align} 
where $|V_{ud}|$ is the first entry in the CKM matrix, $G_F$ is the Fermi constant, $m_e$ is the electron mass, $|M^{2\nu}|$ is the $2\nu\beta\beta$ nuclear matrix element, $t_i$ is the kinetic energy of the $i$th electron in units of electron mass, $F(Z,t_i)$ is the Fermi function describing the Coulomb force between the electron and nucleus taking into account the nuclear size, $\textbf{\textit{p}}_i$ is the momentum of the the $i$th electron, and $\hat{\omega}_0 = Q-t_1-t_2$, where $Q$ is the electron sum-spectrum Q-value.  The coefficient $\mathring{a}^{(3)}_{\text{of}}$ is the parameter of interest in this search, related to the time-like component of the countershaded operator coefficient by $\mathring{a}^{(3)}_{\text{of}} = (a^{(3)}_{\text{of}})_{00}/\sqrt{4\pi}$.

\begin{figure}[b]
\includegraphics[width=0.5\textwidth]{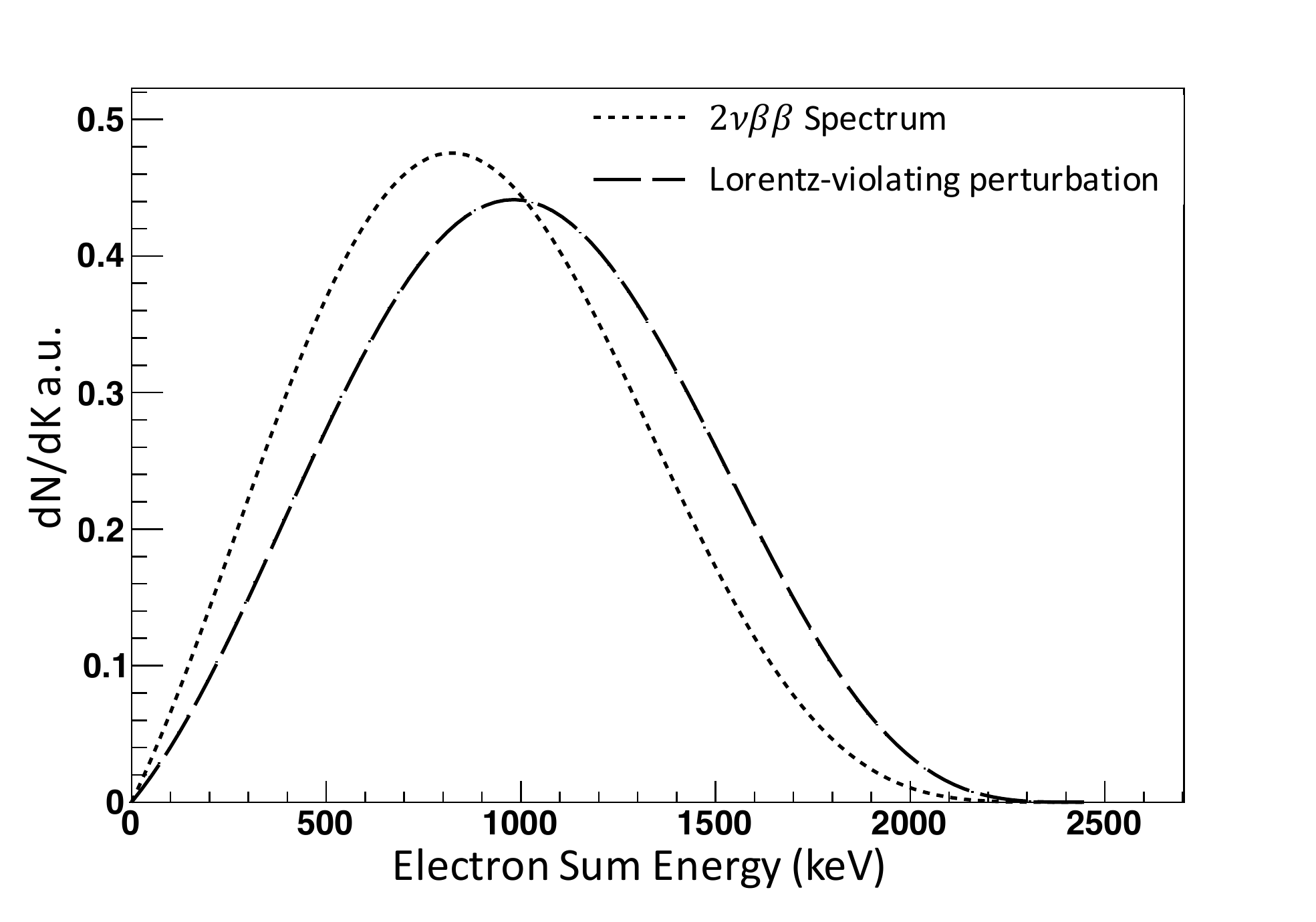}
\caption{\label{fig:2nbb_LVbb} The electron sum spectra of the standard $2\nu\beta\beta$ process (dotted line) compared with the perturbation due to Lorentz-violating effects, LV$\beta\beta$ (dashed line).  Both spectra are normalized to one.}
\end{figure}

This decay rate can be separated into two distinct parts; the quintic term ($\hat{\omega}^5_0$) corresponds to the standard $2\nu\beta\beta$ process \cite{GoeppertMayer:1935qp}, and the quartic term ($\hat{\omega}^4_0$) corresponds to the perturbation of the $2\nu\beta\beta$ spectrum due to the coupling of neutrinos to the Lorentz-violating operator (LV$\beta\beta$).  Precision measurements of the $2\nu\beta\beta$ spectrum require $|\mathring{a}^{(3)}_{\text{of}}| \ll 1$ \cite{Albert:2013gpz}, so the total decay rate can be expressed as an addition of the two separate rates through a perturbation,
\begin{align} \label{eq:DecayRateAdd}
\Gamma = \Gamma_0 + \delta \Gamma
\end{align} 
where $\Gamma_0$ is the decay rate from the standard $2\nu\beta\beta$ spectrum, and $\delta \Gamma$ is the decay rate from the LV$\beta\beta$ perturbation.  The separate decay rates are related to the nuclear matrix element and phase space factors in Eqs. (\ref{eq:StandardDecay}) and (\ref{eq:LVDecay}), where the phase space factor can also be expressed as two separate components through a perturbation, $G^{2\nu} = G^{2\nu}_0 + \delta G^{2\nu}$.  The spectral shapes for both parts, normalized to one, are shown in Figure \ref{fig:2nbb_LVbb}. 
\begin{align}  
\label{eq:StandardDecay}    \Gamma_0 = g^4_Am^2_e|M^{2\nu}|^2G^{2\nu}_0   \\ 
\label{eq:LVDecay}   \delta \Gamma = g^4_Am^2_e|M^{2\nu}|^2\delta G^{2\nu} 
\end{align} 
\begin{align}
 \nonumber
 \label{eq:PhaseSpaceIntegral_2nu} G^{2\nu}_0 = \mathbb{C} \int _0^{Q} dt_1 F(Z, t_1)\sqrt{t_1 (t_1 + 2)} (t_1 + 1)\times  \\ 
\int _0 ^{Q - t_1} dt_2 F(Z, t_2) \sqrt{t_2 (t_2 + 2)} (t_2 + 1) (Q - t_1 - t_2)^5, \\
 \nonumber
 \label{eq:PhaseSpaceIntegral} \delta G^{2\nu} = 10 \mathring{a}^{(3)}_{\text{of}} \mathbb{C} \int _0^{Q} dt_1 F(Z, t_1)\sqrt{t_1 (t_1 + 2)} (t_1 + 1)\times  \\ 
\int _0 ^{Q - t_1} dt_2 F(Z, t_2) \sqrt{t_2 (t_2 + 2)} (t_2 + 1) (Q - t_1 - t_2)^4
\end{align} 

Eqs.~(\ref{eq:PhaseSpaceIntegral_2nu}) and (\ref{eq:PhaseSpaceIntegral}) show the expressions for the two components of the phase-space factor with $\mathbb{C} = G_F^4 |V_{ud}|^4 m_e^9 / 240\pi^7$, with all energy variables in terms of $m_e$.  The SME coefficient $\mathring{a}^{(3)}_{\text{of}}$ only affects the phase-space factor perturbation $\delta G^{2\nu}$.  This can be related to an effective decay rate of the Lorentz-violating perturbation to the $2\nu\beta\beta$ spectrum, with the positive value of $\mathring{a}^{(3)}_{\text{of}}$ searched for with a similar method to a recent search by EXO-200 for Majoron modes of double beta decay \cite{Albert:2014fya}.  The negative value is searched for by subtracting the Lorentz-violating shape from the $2\nu\beta\beta$ spectrum.



The $2\nu\beta\beta$ nuclear matrix element depends on the transition between nuclear states, which is independent of the effects of $\mathring{a}^{(3)}_{\text{of}}$.  Its value is the same for Eqs. (\ref{eq:StandardDecay}) and (\ref{eq:LVDecay}), calculated from the magnitude of the $2\nu\beta\beta$ spectrum.  For the upper and lower bounds of LV$\beta\beta$, Eq.~(\ref{eq:StandardDecay}) is used to calculate $|M^{2\nu}|$ from the number of $2\nu\beta\beta$ counts, which is then used to calculate the limits on $\mathring{a}^{(3)}_{\text{of}}$.


\section{Search Strategy}

This analysis uses the same event reconstruction and fitting techniques as described in detail in previous publications \cite{Albert:2013gpz, Albert:2014awa, Albert:2014fya}.  The same data set is also used, consisting of a total exposure of 100 kg$\cdotp$yr acquired between September 2011 and September 2013.  Probability density functions (PDFs) for the $2\nu\beta\beta$ and LV$\beta\beta$ signals and expected backgrounds are produced using the Geant4-based \cite{Geant4} EXO-200 simulation software, which is described in detail elsewhere \cite{Albert:2013gpz}.  Both data and PDFs are separated into single-site (SS) and multi-site (MS) spectra according to the number of separate charge clusters observed.  A simultaneous fit to the SS and MS spectra is performed to constrain both the $\beta$-like signal events, which are primarily SS, and the $\gamma$-like backgrounds, which are primarily MS.  The fraction of SS vs. MS events for each PDF is constrained based on calibration studies with external $\gamma$ sources.  The observables from the fit are event energy, which is calculated from a linear combination of charge and scintillation channels that optimizes the energy resolution of the $^{208}$Tl line near the $^{136}$Xe Q-value (1.53\% at 2457.83 $\pm$ 0.37 keV \cite{QvalueRef}), and ``standoff distance'', which is the shortest distance between the various charge depositions and anode plane or reflector surface. 

The energy scale for the EXO-200 detector is established with external $\gamma$ sources, which may differ from the energy scale of beta events.  A fit parameter called the ``$\beta$-scale'' ($B$) is defined to describe this potential difference, relating the $\gamma$-like and $\beta$-like energies with $\text{E}_{\beta} = B \text{E}_{\gamma}$, where E$_{\beta}$ (E$_{\gamma}$) is the measured energy deposition from $\beta$'s ($\gamma$'s).  This parameter has been well constrained in previous analyses \cite{Albert:2013gpz,Albert:2014awa}, but due to the similarity in shape between the $2\nu\beta\beta$ and LV$\beta\beta$ spectra the methods previously used are not applicable for this analysis.  The best fit $\beta$-scale is consistent with 1, but the floating $\beta$-scale parameter provides the dominant systematic error for this analysis.

The analysis region for this search is between 980 and 9800 keV.  PDFs for expected backgrounds and the $2\nu\beta\beta$ and LV$\beta\beta$ signal functions are fit to the selected data by minimizing the negative-log likelihood function.  A profile likelihood scan over the number of LV$\beta\beta$ integral counts added to or subtracted from the standard $2\nu\beta\beta$ spectrum is used to obtain limits at the 90\% confidence level (CL).


Several studies were performed on the background model to obtain gaussian constraints on systematic uncertainties for the negative-log likelihood fit.  The radon in the active liquid xenon has been constrained to within 10\% of the activity determined from independent measurements.  In addition, the single-site fraction (SS/[SS+MS]) of each PDF is constrained to within 4\% of the mean calculated value, with the error arising from the largest difference in a binned comparison between source calibration data and Monte Carlo.  An intensive study of the cosmogenic neutron capture gammas was performed \cite{EXO200::2015wtc}, and PDFs from neutron capture gammas on the surrounding materials are constrained together with a 20\% error.  The uncertainty on the background model was estimated by varying the locations of the main background sources and conservatively using the largest variation as an overall background normalization error (20\%).  This error includes the effects of perturbations to the $2\nu\beta\beta$ spectrum due to corrections to the Fermi function arising from the finite nuclear size [39, 40] and corrections due to weak magnetism [41]. In the case of a differing Fermi function, the $2\nu\beta\beta$ PDF integrals differed by 1.5\%, and in the case of including weak magnetism effects the difference was $<$ 1\%.  An overall normalization error of 8.6\% is included, as well as a normalization term specifically for the LV$\beta\beta$ PDF of 29\% to account for differences in the shape of the signal between data and Monte Carlo.  More information about the constraints can be found in \cite{Albert:2014awa}.

\section{Results}

\begin{figure}[b]
\includegraphics[width=0.5\textwidth]{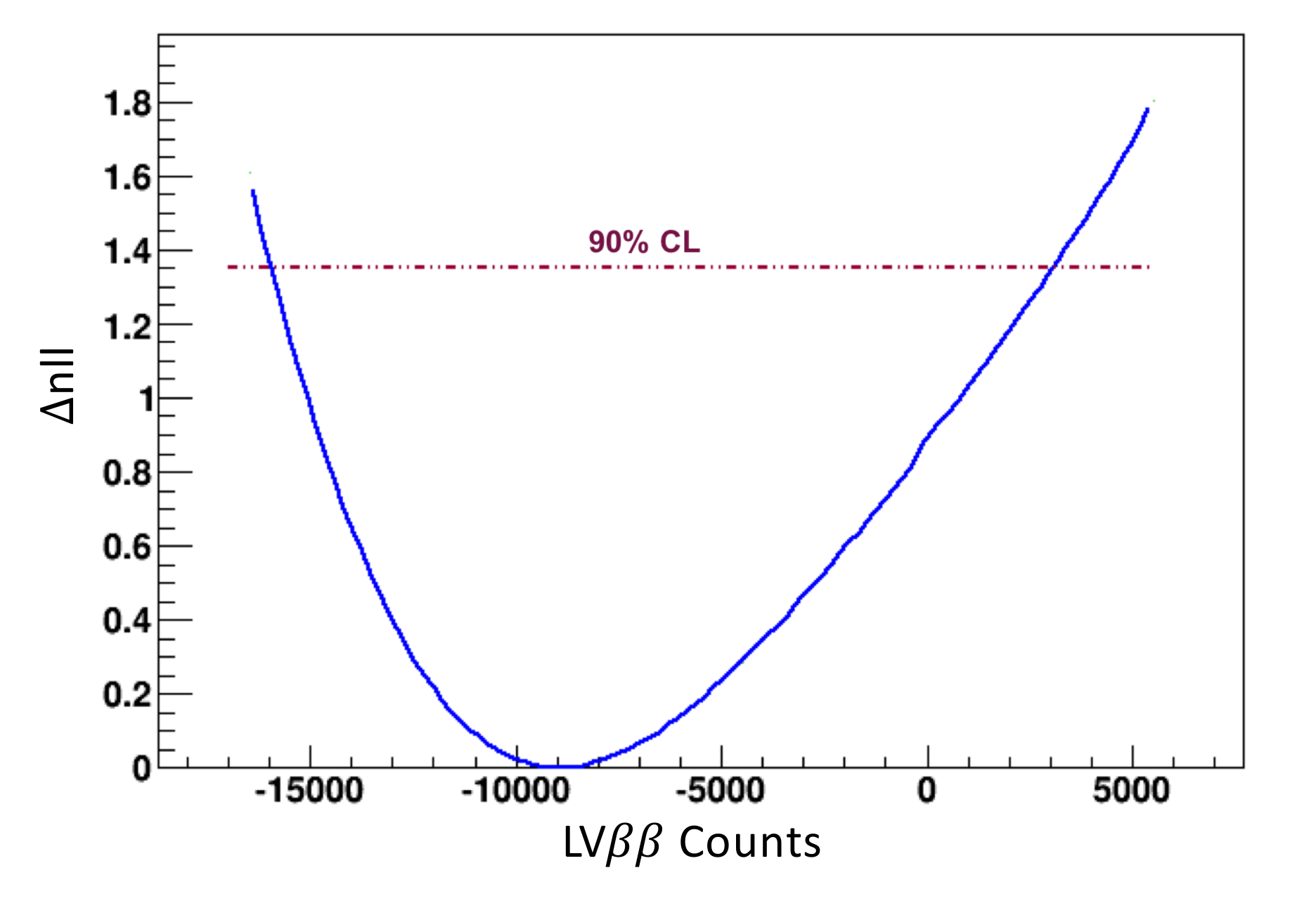}
\caption{\label{fig:Profile_FloatBetaScale} The profile likelihood scan over LV$\beta\beta$ counts is shown by the solid line, with the 90\% CL highlighted by the dashed line, assuming the validity of Wilks' theorem \cite{Wilks:1938dza, SDA}.}
\end{figure}

\begin{figure*}[htbp]
\includegraphics[width=0.995\textwidth]{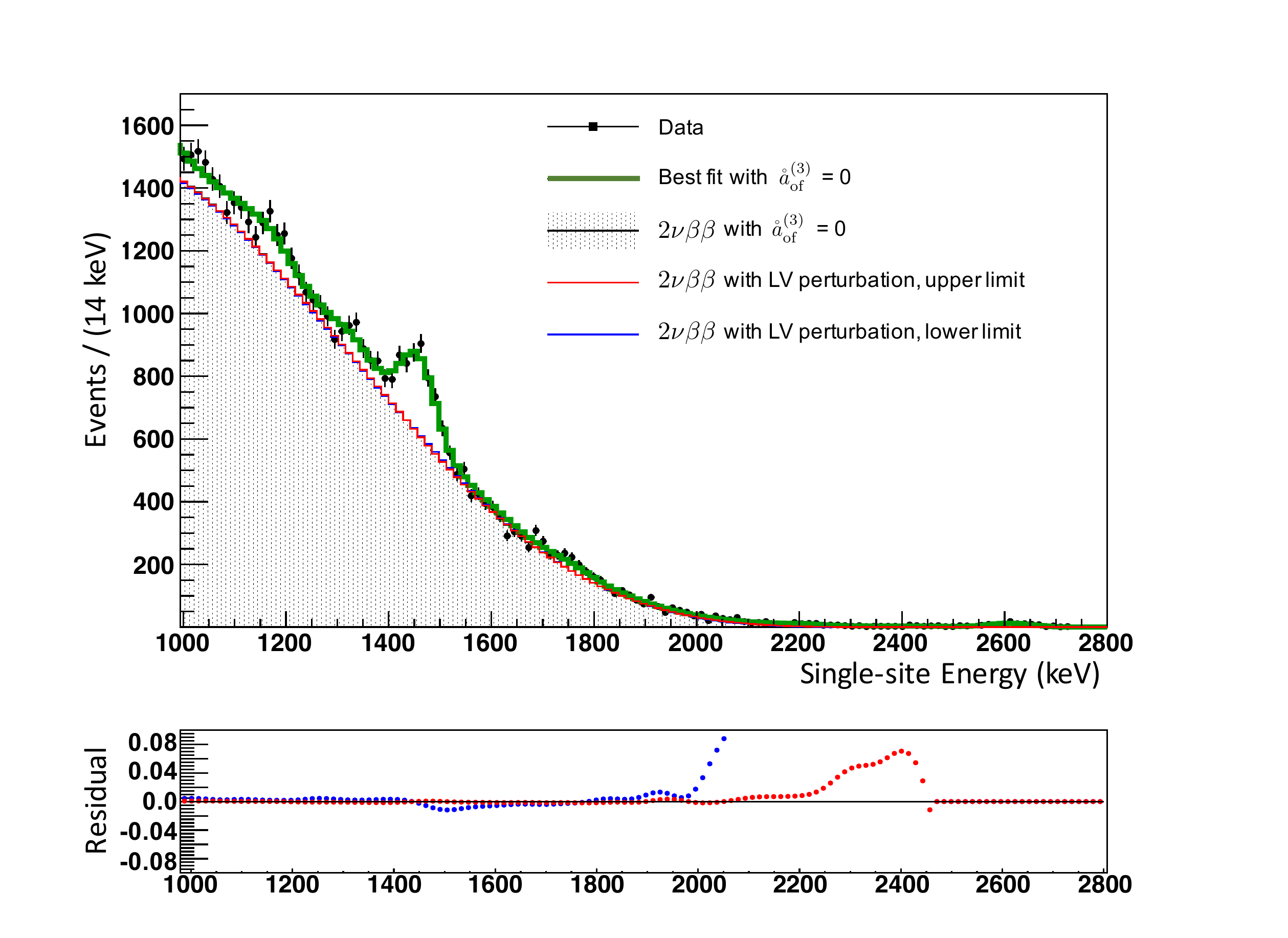}
\caption{\label{fig:Result} The SS data for the case of zero perturbation due to Lorentz-violation are shown with the best overall fit in the energy spectrum, with the best fit of the perturbed $2\nu\beta\beta$ spectra at the upper and lower 90\% CL bounds of the fit for Lorentz-violation superimposed onto the spectrum.  These perturbed $2\nu\beta\beta$ spectra are also shifted in energy according to the best fit $\beta$-scale of the fits.  The fractional residual difference in the total number of counts per bin between the $2\nu\beta\beta$ spectrum in the case of no Lorentz-violation and the upper and lower bound cases is shown on the lower figure, highlighting the difference in the spectra near the $2\nu\beta\beta$ Q-value.  The lower limit residual diverges from the range of the plot at energies near the endpoint due to the shift in the $\beta$-scale between the two compared spectra.}
\end{figure*}

A profile likelihood scan was performed over both positive and negative contributions of LV$\beta\beta$ counts, altering the standard $2\nu\beta\beta$ with both positive and negative values of $\mathring{a}^{(3)}_{\text{of}}$ as shown in Figure \ref{fig:Profile_FloatBetaScale}.  For each profile point in LV$\beta\beta$ counts, a profile likelihood scan was performed over the $\beta$-scale to find the best fit value for that point, broadening the profile by about a factor of 10 from a profile with a fixed $\beta$-scale value.  The scan shows a non-zero best fit value, but is consistent with zero at the 90\% CL.  Using Eqns. (\ref{eq:DecayRateAdd} - \ref{eq:PhaseSpaceIntegral}), the number of LV$\beta\beta$ counts at the 90\% CL was converted into limits on the parameter of interest of $-2.65 \times 10^{-5 } \; \textrm{GeV} < \mathring{a}^{(3)}_{\text{of}} < 7.60 \times 10^{-6} \; \textrm{GeV}$.  The perturbed $2\nu\beta\beta$ spectra with $\mathring{a}^{(3)}_{\text{of}}$ at the 90\% CL limits are shown in Figure \ref{fig:Result}.

The results were checked against the presence in the LXe or detector of exotic isotopes with gamma lines interfering with the analysis region.  The difference in the number of LV$\beta\beta$ counts with included PDFs for $^{88}$Y and $^{110\textrm{m}}$Ag was found to be $\ll 1\sigma$.

Another group has previously calculated a limit on the parameter of interest of $|(a^{(3)}_{\text{of}})_{00}|<2 \times 10^{-8}$ GeV by performing an outside analysis on the endpoint of the Mainz tritium beta decay data \cite{KM_NeutrinosAndLV}.  However, the limit presented by EXO-200 is the result of the first search for this parameter that fully accounts for experimental backgrounds and detector-related systematic uncertainties.  The application of the techniques described in this work to the substantially larger data sets available with single beta decay sources \cite{LVSingleBeta} may allow even further improvements in sensitivity.


In conclusion, we report on the first experimental search in double beta decay for the the isotropic component of the coefficient describing the momentum-independent and oscillation-free operator coupling to neutrinos in the Standard-Model Extension.  We detect no significant signal from studying the potential shape deviation from the standard $2\nu\beta\beta$ spectrum and set limits on the magnitude of this coefficient.  Future work to independently constrain the beta energy scale could allow substantial improvement in sensitivity to this parameter with EXO-200.  

\begin{acknowledgments}

The collaboration gratefully acknowledges the KARMEN collaboration for supplying the cosmic-ray veto detectors, the WIPP for their hospitality, and the support from the Indiana University Center for Spacetime Symmetries (IUCSS). J.~S.~D. was supported in part by DFG (KL 1103/4-1).  EXO-200 is supported by DOE and NSF in the United States, NSERC in Canada, SNF in Switzerland, IBS in Korea, RFBR-14-02-00675 in Russia, DFG Cluster of Excellence ``Universe'' in Germany, and CAS-IHEP Fund and ISTCP (2015DFG02000) in China. EXO-200 data analysis and simulation uses resources of the National Energy Research Scientific Computing Center (NERSC), which is supported by the Office of Science of the U.S. Department of Energy under Contract No. DE-AC02-05CH11231.

\end{acknowledgments}

\end{document}